\input harvmac

\def\p{\partial}

\Title{hep-th/0010058}{\vbox{\centerline{Note on Noncommutative Tachyon in Matrix Models}}}
\centerline{ Miao Li }
\centerline{\it Institute for Theoretical Physics}
\centerline{\it Academia Sinica, Beijing}
\centerline{And} 

\centerline{\it Department of Physics}
\centerline{\it National Taiwan University}
\centerline{\it Taipei 106, Taiwan} 
\centerline{\tt mli@phys.ntu.edu.tw}

\vskip1.5cm

The solution representing a brane-anti-brane system in matrix models
breaks the usual matrix spacetime symmetry. We show that the spacetime
symmetry on the branes is not breaking, rather appears as a combination
of the matrix spacetime transformation and a gauge transformation.
As a result, the tachyon field, itself an off-diagonal entry in
longitudinal matrices, transforms nontrivially under rotations, decomposing
into tensors of different ranks. We also show that the tachyon field
can never be gauged away, and conjecture
that this field is related to the usual complex scalar tachyon by
a field redefinition. We also briefly discuss tachyon condensation. 

\vskip0.5cm
\Date{October 2000}

\nref\tachy{A. Sen,``SO(32) spinors of type I and other solitons
on brane-antibrane pair," JHEP 9809 (1998) 023, hep-th/9808141;
``Stable non-BPS bound states of BPS D-branes," JHEP 9808 (1998) 010,
hep-th/9805019; ``Tachyon condensation on the brane antibrane system,"
JHEP 9808 (1998) 012, hep-th/9805170; ``BPS D-branes on non-supersymmetric
cycles," JHEP 9812 (1998) 021, hep-th/9812031; ``Descent relations
among bosonic D-branes," Int. J. Mod. Phys. A14 (1999) 4061, hep-th/9902105.}
\nref\kth{E. Witten, ``D-branes and K-theory," JHEP 9812 (1998) 019,
hep-th/9810188; P. Horava, ``Type IIA D-branes, K-theory, and matrix theory,"
Adv. Theor. Math. Phys. 2 (1999) 1373, hep-th/9812135.}
\nref\gms{R. Gopakumar, S. Minwalla and A. Strominger,
``Noncommutative Solitons,''
JHEP {\bf 0005} (2000) 020, hep-th/0003160.}
\nref\dmr{K. Dasgupta, S. Mukhi and G. Rajesh,
``Noncommutative Tachyons,''
JHEP {\bf 0006} (2000) 022, hep-th/0005006.}
\nref\hklm{
J. A. Harvey, P. Kraus, F. Larsen and E. J. Martinec,
``D-branes and Strings as Non-commutative Solitons,''
JHEP {\bf 0007} (2000) 042, hep-th/0005031.}
\nref\ew{E. Witten,
``Noncommutative Tachyons And String Field Theory,''
hep-th/0006071.}
\nref\wit{E. Witten, ``Overview Of K-Theory Applied To Strings,"
hep-th/0007175.}
\nref\sw{N. Seiberg and E. Witten, ``String Theory and Noncommutative Geometry,"
HEP 9909 (1999) 032, hep-th/9908142.}
\nref\bfss{T. Banks, W. Fischler, S. Shenker and L. Susskind, 
``M Theory As A Matrix Model: A Conjecture," Phys.Rev. D55 (1997) 5112,
hep-th/9610043.}
\nref\iktt{N. Ishibashi, H. Kawai, Y. Kitazawa and A. Tsuchiya,
``A Large-N Reduced Model as Superstring," Nucl.Phys. B498 (1997) 467,
hep-th/9612115.}
\nref\miao{M. Li, ``Strings from IIB matrices," Nucl. Phys.
B499 (1997) 149, hep-th/9612222.}
\nref\seib{N. Seiberg, ``A Note on Background Independence 
in Noncommutative Gauge Theories, Matrix Model
and Tachyon Condensation," hep-th/0008013.}
\nref\ha{H. Awata, S. Hirano and Y. Hyakutake, ``Tachyon Condensation 
and Graviton Production in Matrix Theory," hep-th/990215.}
\nref\krs{P. Kraus, A. Rajaraman and S. Shenker, ``Tachyon
Condensation in Noncommutative Gauge Theory," hep-th/0010016.} 
\nref\tatar{R. Tatar, ``A Note on Non-Commutative Field Theory and 
Stability of Brane-Antibrane Systems," hep-th/0009213.}
\nref\gmas{R. Gopakumar, S. Minwalla and  Strominger,
``Symmetry Restoration and Tachyon Condensation in Open String 
Theory,"  hep-th/0007226.}
\nref\sen{A. Sen, ``Some Issues in Non-commutative Tachyon Condensation,"
 hep-th/0009038; ``Uniqueness of Tachyonic Solitons,"
hep-th/0009090.}
\nref\piljin{P. Yi, ``Membranes from Five-Branes and Fundamental Strings from Dp 
Branes," Nucl.Phys. B550 (1999) 214,  hep-th/9901159;
O. Bergman, K. Hori and P. Yi, ``Confinement on the Brane,"  Nucl.Phys. B580 (2000) 289,
hep-th/0002223.}
\nref\ashoke{A. Sen, ``Supersymmetric World-volume Action for Non-BPS D-branes,"
JHEP 9910 (1999) 008, hep-th/9909062;
`` Universality of the Tachyon Potential," 
JHEP 9912 (1999) 027, hep-th/9911116.}
\nref\chl{C.S. Chu, P.M. Ho and M. Li, ``Matrix Theory in a Constant 
C Field Background," Nucl.Phys. B574 (2000) 275, hep-th/9911153.}

\newsec{Introduction}

Tachyon condensation has been one of the focus points of recent research 
in string theory \refs{\tachy, \kth}.  One apparent motivation in studying 
this  issue is to understand the dynamic
process of the brane-anti-brane annihilation. More profoundly, revelation 
of elements in this problem will shed new lights on the deep connection 
between the open string sector and the closed string sector, and on the old 
problem of background independence in string theory. The tachyon condensation 
problem has been proven
considerably simplified when a large B field is present \refs{\gms-\ew}, thanks
to some new features of noncommutative field theory \sw.

As conjectured recently by Witten \wit, it may be possible to realize any
 configuration in, for instance the IIB theory, by starting with 
infinitely many D9 and anti-D9
branes. It is also very interesting to ask whether it is possible to 
``derive" the IIB matrix model or some revision of it \refs{\bfss, \iktt}, 
by starting with D9-anti-D9 system and
turning on tachyon condensation. We believe that the answer to this question is
likely yes. The reason for this belief is the following. Consider for instance 
a single D9-brane with a large B field switched on. The open string sector
decouples from the closed string sector in the Seiberg-Witten scaling limit \sw.
If the rank of $B$ is maximal (in the Euclidean spacetime), then one can
rewrite the noncommutative Yang-Mills theory as a 0-dimensional matrix model in
an operator formulation.
Since there is no tachyon in this system, it is not possible to realize other
D-branes and closed strings yet. By adding anti-D9-branes, one would get
a matrix model different from the IIB matrix model: If both D9-branes and
anti-D9-branes are immersed in the same B field background, again one
has a noncommutative field theory, one that is different from the IIB matrix 
model in an operator formulation.

One can study this problem by reversing the above procedure. One starts with
the IIB matrix model and constructs D9-anti-D9 solution. This solution is a 
system quite different from D9-anti-D9 embedded in the same constant B
field background. The reason for this is quite simple. If the D9-brane is 
represented by a solution $[X^i,X^j]=i\theta^{ij}$, then an anti-D9-brane is
represented by a solution $[X^i,X^j]=-i\theta^{ij}$, corresponding to
reversing the orientation of the D9-brane solution by making reflection of 
5 dimensions. However, a D-brane solution in matrix models does not come 
for free, it always carries a background gauge field $F_{ij}$ which is just the
inverse of $\theta^{ij}$.  Indeed for this solution, as first observed 
in \miao, a noncommutative
super Yang-Mills theory directly results from the matrix model. 
For the brane-anti-brane solution, there are different 
background gauge fields on D9-branes and anti-D9-branes. It is not
known how to write down a simple field theory with an associative 
algebra generalizing the star product. On the D9-branes, the star product
is defined with the noncommutative parameters $\theta^{ij}$, while on
the anti-D9-branes, the star product is defined with $-\theta^{ij}$.
If one naively generalizes the product to include both, one will not
get an associative algebra.

As we mentioned before, if one starts with a background in which the gauge
field strengths on all branes are the same, then one has the usual noncommutative
field theory. Other different field strengths can be achieved by adding
perturbations. To simplify the situation, we shall focus on two coordinates,
and the resulting theory can either be regarded as D-string anti-D-string
in the IIB matrix model, or D2-brane anti-D2-brane in matrix theory. 
The matrix model solution of this configuration explicitly realizes the above
idea. As a result, the solution breaks the usual matrix model rotational
symmetry. This can be seen in two different ways. The simpler one is
to simply take a look at the solution $X^1=\hbox{diag}(x^1,x^1)$, $X^2
=\hbox{diag}(x^2, -x^2)$ with $[x^1,x^2]=i\theta$. The first entries 
represent the D-string, and the second entries represent the anti-D-string.
This solution is not invariant under the rotation $\delta X^i= \epsilon 
\epsilon^{ij}X^j$. The second way to see this is to write $X^2=\hbox{diag}
(x^2,x^2) +\hbox{diag} (0,-2x^2)$. The second term can be regarded as a 
perturbation in the gauge field on the second string and this field depends
explicitly on coordinates. This is a puzzle, since we believe that even
though the field strengths on different strings are different, the whole
system is invariant under a rotation. We will see in the next section
that the rotational symmetry is still present in the solution. This symmetry
is a mixture of the matrix rotation and a gauge transformation.
In sect.3, we shall see that with the presence of gauge fields, the mixture
is more complicated. This is not surprising, since gauge fields can be
understood as longitudinal oscillations of branes, thus it is intimately
related to spacetime symmetry.

The scalar nature of the tachyon presents another puzzle. We expect this mode
to arise from the off-diagonal entries of $X^i$, and these entries carry
longitudinal spacetime indices. This puzzle again is related
to the lacking of matrix rotational symmetry. We will in sect.4 see that
one of the complex off-diagonal entry can be gauged away, the remaining
off-diagonal entry is tachyonic. Under the rotation about a point, 
the tachyon can be divided into modes with different tensor structures.
We attribute this unusual feature to the bad representation of brane-anti-brane
system in matrix models. We believe that if one can write down an
associative algebra without breaking the matrix rotational symmetry,
the scalar nature of the tachyon will become apparent. We briefly discuss
tachyon condensation in the final section.

A paper discussing related issues appeared very recently \krs.

\newsec{Rotational Invariance}

In the Euclidean signature, the brane-anti-brane solution can be
represented as
\eqn\confi{X^1=\hbox{diag} (x^1, x^1),\quad X^2 =\hbox{diag} (x^2, -x^2).}
Here $x^1,x^2$ are noncommutative coordinates satisfying
\eqn\commt{ [x^1,x^2]=i\theta.}
Apparently, the brane-anti-brane system still respects the rotational
invariance on the $(x^1,x^2)$ plane, this is a consequence of the
first quantized strings for them the world-sheet action is invariant
under rotations. For an infinitesimal rotation,
$\delta x^1=-\epsilon x^2, \delta x^2=\epsilon x^1$, we expect
the matrices $X^1, X^2$ transform in the same way. This is not the case
for the ansatz \confi, since $X^1$ is proportional to the identity matrix
while $X^2$ is proportional to $\sigma_3$. Thus we can hardly obtain
a manifestly rotationally invariant action or Hamiltonian starting with
ansatz \confi.

In matrix models, in addition to translational and rotational symmetry,
the only other symmetry readily available is gauge symmetry. Applying a
rotational transformation directly to $X^1, X^2$ would bring us
away from the ansatz \confi. To come back to \confi, one may try to
apply a gauge transformation. It is easy to guess what kind of gauge
parameter is needed. Since the first diagonal entries  transform
correctly under the matrix spacetime transformation, thus the gauge parameter
has a vanishing first diagonal entry. Assume $g=\hbox{diag} (0, f)$,
where $f$ is a function of $x^{1,2}$, the gauge transformation takes 
the form $\delta_g X^i=i[g, X^i]$.
Now apply the rotational transformation
$$\delta_\epsilon X^1=\hbox{diag} (-\epsilon x^2, \epsilon x^2),\quad
\delta_\epsilon X^2=\hbox{diag} (\epsilon x^1, \epsilon x^1) .$$
Applying further a gauge transformation, in order to keep the ansatz 
\confi\ invariant, we need
\eqn\cond{\eqalign{\epsilon x^2 +i[f,x^1]&=-\epsilon x^2,\cr
\epsilon x^1-i[f,x^2]&=-\epsilon x^1.}}
The apparent solution to the above conditions is
\eqn\solu{f=-{\epsilon\over\theta}[ (x^1)^2+(x^2)^2].}
The corrected rotational transformation therefore is
\eqn\ctran{\eqalign{\delta X^1&=(\delta_\epsilon+\delta_g)X^1
=\hbox{diag} (-\epsilon x^2, -\epsilon x^2),\cr
\delta X^2&=(\delta_\epsilon+\delta_g)X^2=\hbox{diag} (\epsilon x^1,
-\epsilon x^1).}}
This is of course a symmetry since both terms on the R.H.S. are
symmetries in the matrix model.

Note that the gauge transformation with parameter \solu\ is rather
singular in the usual sense, for $f$ diverges for large $r^2$. 
This is not surprising however, since the rotational transformation
itself diverges too. 

Similarly, ansatz \confi\ does not conform with the conventional translational
transformation $\delta X^i =a^i\times 1_2$, where $1_2$ is the identity
matrix. To preserve the form \confi, the conventional transform is
to be accompanied by a gauge transformation with a gauge parameter
\eqn\tgg{g=\hbox{diag} (0, -2(a^2/\theta) x^1).}

\newsec{The Gauge Fields}

The same issue of maintaining rotational transformation remains when
we perturb from the constant background \confi. For instance,
consider switching on the $U(1)$ gauge field corresponding to
the center of mass degree of freedom. In case of two coincident
D-branes, the ansatz in matrix models is \miao
\eqn\tcoin{X^i= (x^i+\theta\epsilon^{ij}A_j)1_2.}
The commutator $[X^1,X^2]$ is given by 
\eqn\ccomm{[X^1,X^2]=i\theta (1+\theta F_{12})1_2,}
where $F_{12}$ is the noncommutative field strength
\eqn\fstr{F_{12}=\p_1 A_2-\p_2 A_1-i[A_1,A_2],}
and we implicitly used the star product defined by the
noncommutative parameter $\theta$. Thus, the
action is proportional to
\eqn\actp{\int {d^2x\over 2\pi\theta}\theta^2(1+\theta F_{12})^2.}
As noticed in \seib, $\theta^2$ is to be interpreted
as the open string metric factor $G^{ij}$. The
integral factor is simply the phase space factor
corresponding to the matrix trace.

The correct ansatz generalizing
both \confi\ and the usual brane-brane system is
\eqn\cuone{X^1=(x^1+\theta A_2) 1_2,\quad
X^2=(x^2-\theta A_1)\sigma_3.}
The commutator $[X^1,X^2]$ is the same as in \ccomm\ with an
additional factor $\sigma_3$. This does not change
the action.
Again, under the usual rotational transformation $\delta_\epsilon
X^i=-\epsilon \epsilon^{ij}X^j$, the above ansatz is not preserved. 
We expect that
with the coordinates rotation $\delta x^i=-\epsilon\epsilon^{ij}x^j$,
the gauge field $A_i$ transform in the same fashion, namely $\delta A_i
=-\epsilon\epsilon_{ij}A_j$. This together with the ansatz \cuone\ implies
that an additional gauge transformation is again needed:
\eqn\rotat{\delta X^i=\delta_\epsilon X^i+i[g,X^i].}
Again, $g$ is diagonal and assumes the form $g=\hbox{diag} (0, 2\epsilon f)$.
The conditions that $f$ must meet are
\eqn\mcon{\eqalign{i[f,x^1+\theta A_2]&=-(x^1-\theta A_1),\cr
i[f, x^2-\theta A_1]&=x^1+\theta A_2.}}
The commutators in the above can be written as covariant derivatives,
so formally these conditions can be written as 
\eqn\fcon{D_if=-i\epsilon^{ij}D_j.}

With the presence of the other $U(1)$ gauge field $B_i$
corresponding to the difference of gauge fields on
the two branes, the correct ansatz is
\eqn\fansa{\eqalign{X^1&=(x^1+\theta A_2) 1_2 +\theta B_2\sigma_3,\cr
X^2&=(x^2-\theta A_1)\sigma_3 -\theta B_1 1_2.}}
To see that this is indeed the correct ansatz, we write
the above formulas in the component form
\eqn\comfo{\eqalign{X^1&=\hbox{diag}(x^1+\theta (A_2+B_2), x^1+
\theta (A_1-B_1)),\cr
X^2&=\hbox{diag}(x^2-\theta (A_1+B_1), -[x^2-\theta (A_1-B_1)]).}}
Indeed we see that $A_i+B_i$ is the gauge field living on
the brane, and $A_i-B_i$ is the gauge field living
on the anti-brane.
Again it is the second entries on the anti-brane that break the
usual matrix rotational symmetry. To restore the
coordinates rotational symmetry, we need to form
a gauge transformation with a gauge parameter of the
form $g=\hbox{diag}(0, 2\epsilon f)$ and $f$ satisfies
\eqn\fcon{\eqalign{i[f, x^1+\theta (A_2-B_2)]&=-(x^2-\theta 
(A_1-B_1),\cr
i[f,x^2-\theta (A_1-B_1)]&=x^1+\theta (A_2-B_2).}}
Or more compactly
\eqn\compc{D_i(A-B)f=-i\epsilon^{ij}D_j(A-B),}
where $D_i(A-B)$ is the covariant derivative defined
with respect to the gauge field $A-B$.

The commutator $[X^1,X^2]$ assumes a little more
complicated form than for the brane-brane system
\eqn\fcomm{[X^1,X^2]=i\theta \left[ (F_{12}+\theta^{-1})
\sigma_3 +\tilde{F}_{12} 1_2\right],}
where it is not surprising to see the
term $\theta^{-1}$ since the $B$ field is $-\theta^{-1}$.
The two $U(1)$ field strengths are defined by
\eqn\twostr{\eqalign{F_{12}&=\p_1A_2-\p_2A_1-i[A_1,A_2]-i[B_1,B_2],\cr
\tilde{F}_{12}&=D_1(A)B_2-D_2(A)B_1,}}
where the covariant derivatives $D_i(A)$ are defined against
the gauge field $A_i$. It is easy to understand why
these covariant derivatives appear in $\tilde{F}_{12}$,
since everything is charged adjointly with respect to the center
of mass $U(1)$ gauge field. However, it is rather unusual
that the gauge field $B_i$ contributes to $F_{12}$, the
field strength of the center of mass degree of freedom.
The cause of this is the asymmetric fashion in which we are
dealing with the brane and the anti-brane.

In a noncommutative field theory it is often convenient
to work with the creation and annihilation coordinates
\eqn\cran{a={1\over\sqrt{2\theta}} (x^1+ix^2),\quad
a^+={1\over\sqrt{2\theta}}(x^1-ix^2)}
satisfying $[a,a^+]=1$. These are the analogue of
complex coordinates $z=x^1+ix^2$, $\bar{z}=x^1-ix^2$.
The gauge fields we introduced can also be written in
the complex form
\eqn\gacom{A_z=A_1-iA_2, \quad A_{\bar{z}}=A_1+iA_2.}
It is straightforward to see that $A_z$ always appears
together with $a^+$ which can be regarded as $\p_z$, and
$A_{\bar{z}}$ always appears with $a$ which can be written
as $\p_{\bar{z}}$.

\newsec{The Tachyon}

We expect the tachyon to arise as the off-diagonal entries of matrices 
$X^1, X^2$. Modes from other matrices remain massless. In the 
brane-anti-brane system, the tachyon field is a complex scalar. However, 
one naively expects that any field  arising from $X^i$ carries a 
spacetime index tangent to the branes. The resolution of this puzzle is 
connected to the fact that the background \confi\ breaks the usual 
spacetime symmetry in the matrix model, and the conventional world-volume
spacetime symmetry, as explained in the previous two sections, is a 
combination of the matrix model spacetime symmetry and gauge symmetry.

We work again with the harmonic representation. The gauge parameter 
associated with rotation generator can be written as
\eqn\hgt{g=\left(\matrix{0&0\cr 0 &-\epsilon (2N+1)}
\right),}
where $N=a^+a$ is the number operator, $\epsilon$ is the infinitesimal
rotation angle. As explained before, the rotation around point $z
=x^1+ix^2=0$ is generated by
\eqn\rotg{\delta X^i=-\epsilon\epsilon^{ij}X^j+i[g,X^i].}
It is convenient to work with the complex matrices
\eqn\comm{Z={1\over\sqrt{2}}(X^1+iX^2),\quad
\bar{Z}={1\over\sqrt{2}}(X^1-iX^2).}
Since $\bar{Z}$ is the Hermitian conjugate of $Z$, it is enough to
work with $Z$. In the absence of the gauge fields, 
\eqn\tachy{ Z=\sqrt{\theta}\left(\matrix{a & \tilde{T}\cr T&a^+}\right).}
Under the rotational transformation \rotg, $Z$ transforms as
\eqn\ztrans{\delta Z=i\epsilon Z-i\epsilon\sqrt{\theta}\left(\matrix{
0&-\tilde{T}(2N+1)\cr (2N+1)T&2a^+}\right).}
We read from above that $\delta a=i\epsilon a$ and $\delta a^+
=-i\epsilon a^+$, meaning that $a$ transforms as $z$ and $a^+$
as $\bar{z}$. The off-diagonal fields transform according to
\eqn\offt{\delta T =-2i\epsilon NT,\quad \delta\tilde{T}
=2i\epsilon \tilde{T}(N+1).}
These formulas say that under the rotation, both $T$ and
$\tilde{T}$ transform quite nontrivially, and that one can
not simply classify them by a tensor of a fixed rank.

$T$ and $\tilde{T}$ are both operator-valued, so it is convenient
to expand them in terms of the basis $|m\rangle\langle n|$
where $N|m\rangle =m |m\rangle$.
\eqn\expan{T=\sum T_{mn}|m\rangle\langle n|,\quad
\tilde{T}=\sum\tilde{T}_{mn}|m\rangle\langle n|.}
Thus under a rotation, $T_{mn}$ transforms as $\bar{z}^{2m}$
and $\tilde{T}_{mn}$ as $z^{2(n+1)}$. Since $m, n\ge 0$, one
can never get a scalar from $\tilde{T}$, while $T_{0n}$
are invariant under the rotation. 

For a fixed $m$, one can write 
\eqn\oprp{\sum_n T_{mn}\langle n|=\langle 0|T_m}
with an operator $T_m$, so that
\eqn\opprp{T=\sum |m\rangle\langle 0|T_m.}
Now $T_m$ transforms as $\bar{z}^{2m}$. In particular,
$T_0$ is a scalar under the rotation. Similarly,
\eqn\oppt{\tilde{T}=\sum \tilde{T}_m |0\rangle\langle m|,}
and $\tilde{T}_m$ transforms as $z^{2(m+1)}$. 

Although $T_0$ is invariant under rotation about $z=0$, it is
not translationally invariant. This is not a surprise, as
$T_0$ itself is not a fully-fledged two dimensional
field. The state $|0\rangle$ can be regarded as localized
at $z=0$. One can construct a similar state localized at
another point $z_0$ satisfying
\eqn\anlo{a|z_0\rangle ={1\over\sqrt{2\theta}}z_0|z_0\rangle.}
This state can be expressed as a coherent state
\eqn\cohs{|z_0\rangle =e^{{z_0\over\sqrt{2\theta}}a^+}|0\rangle.}
In a sense, $a^+$ can be regarded as the generator of
a translation in the $z$ direction in the noncommutative plane.
This observation agrees well with the gauge parameter
\tgg, since $x^1\sim a+a^+$. Just like $|0\rangle$,
$|z_0\rangle$ is not well-localized in the $\bar{z}$ direction,
as the plane is noncommutative.
Now a tachyon mode
\eqn\ztach{T= |z_0\rangle\langle 0|T}
is invariant under the rotation about the point $z_0$.
The corresponding gauge parameter similar to \hgt\
can be expressed in terms new creation and annihilation
operators $a\rightarrow a-z_0/\sqrt{2\theta}$, $a^+
\rightarrow a^+-\bar{z}_0/\sqrt{2\theta}$.

The action (of strings) or the Hamiltonian (of D2-branes) is
proportional to $[Z,\bar{Z}]^2$:
\eqn\tact{\eqalign{{1\over\theta^2}[Z,\bar{Z}]^2&=(1-T^+T+\tilde{T}\tilde{T}^+)^2+
(1-TT^++\tilde{T}^+\tilde{T})^2\cr
&+2|Ta^+-aT+a^+\tilde{T}^+
-\tilde{T}^+a|^2,}}
where the first two terms on the R.H.S. tell us that $T$ is
the tachyon while $\tilde{T}$ is massive. That $T$ is tachyonic
was first noticed in \ha. The massive mode $\tilde{T}$ should
not be a propagating mode in the brane-anti-brane system.
We now show that indeed this mode can be gauged away.

Although in the brane-anti-brane system, the gauge symmetry
is $U(1)\times U(1)$, the full gauge symmetry in the matrix
model is still $U(2)$, and $T$ and $\tilde{T}$ can be regarded
as the corresponding off-diagonal gauge fields. Since the
off-diagonal gauge parameter is a complex field, one can gauge
away one of them. One would guess that one can gauge away
any of them, or any linear combination of them, as in a
brane-brane system. This is not the case here. We can gauge
away only $\tilde{T}$, the massive mode. To see this, let us
apply the gauge transformation with the gauge parameter
\eqn\gaup{g=\left(\matrix{0&f^+\cr f&0}\right)}
to $Z$, and we find
\eqn\gaut{\eqalign{\delta T &=i(fa-a^+f)\cr
\delta\tilde{T}&=i(f^+a^+-af^+).}}
Similar to \opprp\ and \oppt, we expand
\eqn\gexp{\eqalign{f=&\sum |m\rangle\langle 0|f_m \cr
f^+=&\sum f_m^+|0\rangle\langle m|,}}
where $f_m$, $f_m^+$ are operators.
Using 
$$a|m\rangle =\sqrt{m}|m-1\rangle, \quad
a^+|m\rangle =\sqrt{m+1}|m+1\rangle$$
we find, in terms of the components
\eqn\compex{\eqalign{\delta\tilde{T}_m|0\rangle &=i(\sqrt{m+1}f^+_{m+1}
-af_m^+)|0\rangle \cr
\langle 0|\delta T_m&=i\langle 0|(f_ma -\sqrt{m}f_{m-1}).}}
For an arbitrary $\delta\tilde{T}_m$, it is possible to
satisfy the first equation of \compex, since one simply
needs the following recursion relation
\eqn\recur{\sqrt{m+1}f^+_{m+1}=-i\delta\tilde{T}_m+af_m^+.}
Thus it is possible to gauge away any infinitesimal
$\tilde{T}_m$ by letting $\delta\tilde{T}_m=-\tilde{T}_m$.
On the other hand, for an arbitrary $\delta T_m$, it is
generally impossible to solve the second equation in
\compex. The reason is that in the recursion relation
\eqn\arec{\langle 0|f_m a=\langle 0| (-i\delta T_m
+\sqrt{m}f_{m-1})}
$a$ is not invertible. For instance, for a nonvanishing
c-number valued $\delta T_0$, there is no solution
to the equation
\eqn\zerc{\langle 0|f_0 a=-i\delta T_0\langle 0|.}

Now we choose a gauge in which $\tilde{T}=0$. The residual
gauge symmetry is $U(1)\times U(1)$. This is the
desired gauge symmetry in the brane-anti-brane system. The
action for $T$ becomes quite simple in this gauge
\eqn\actt{{1\over\theta^2} [Z,\bar{Z}]^2=(1-T^+T)^2+
(1-TT^+)^2+2|Ta^+-aT|^2.}
Again, this action is not invariant under the world-volume
rotation if we simply regard $T$ as a scalar. As we
showed before, $T$ actually transforms nontrivially under a
rotation. The breaking-down of an explicit rotational
symmetry when $T$ is treated as a scalar is due to the fact that 
in the matrix model
the background forces us to use the star product which
is natural only for the D-brane, not for the anti-D-brane.
If an algebraic structure symmetric for the pair of brane
and anti-brane exists, the tachyon should appear as a
scalar, and should be related to the tachyon discussed
here by a field redefinition, quite similar to the
Seiberg-Witten field redefinition.

\newsec{Tachyon Condensation}

As discussed in the previous section, the complex tachyon field
does not appear as a scalar in the matrix model solution.
Naturally therefore, we do not expect a uniform tachyon condensation
in this context, as also pointed out in \krs.
First of all, we want to show that there does not exist a tachyon
configuration making $[Z,\bar{Z}]=0$, the absolute minimum,
when gauge fields are set to zero.

From \actt\ the conditions for $[Z,\bar{Z}]=0$ are
\eqn\mini{T^+T=1,\quad a^+=T^{-1}aT.}
The first equation says that $T$ is a unitary operator, and
the second equations says that $a^+$ is similar to $a$.
Taking Hermitian conjugate of this condition we have
$a=T^{-1}a^+T$ or $a^+=TaT^{-1}$. This together with
the second equation in \mini\ implies
\eqn\tscom{[T^2, a]=0.}
Similarly one can derive $[T^2,a^+]=0$. Thus
$T^2$ commutes with all operators constructed of $a$ and
$a^+$ and must be a c-number. Unitarity of $T$ further 
implies that $T^2$ must be a pure phase: $T^2=\exp (2i\phi)$.
In general, $T=\hbox{diag}(\pm \exp(i\phi))$. But this
form can not satisfy $a^+=T^{-1}aT$. We conclude
that there is no absolutely minimal tachyon condensation.

The cause of the absence of the minimal tachyon condensation
is rather obvious, that is in the brane-anti-brane solution
the gauge field strengths on the brane and the anti-brane
have opposite sign. The lacking of a uniform minimal
tachyon condensation is also noticed in \krs. Due to
the nonscalar nature of $T$, we thus expect localized 
minimum at best, coming from solving the equation of motion
derived from the variational principle $\delta\tr [Z,\bar{Z}]^2=0$.
The equation of motion is just
\eqn\eomo{T(1-T^+T)=NT+TN-aTa-a^+Ta^+}
and its Hermitian conjugate. It is not easy to find
a general solution to the above equation. However,
a simple solution making both sides of \eomo\ vanish
exists, it is simply
\eqn\vort{T=c|0\rangle\langle 0|}
with $|c|^2=1$. This is a vortex localized at the origin.
Unlike in the uniform situation studied in \gms, $T\sim
|m\rangle \langle m |$ can not be an exact solution,
since both $aTa$ and $a^+Ta^+$ bring $T$ out of the
projection operator. According to the discussion in
the previous section, this is not surprising, since
these modes are not rotationally invariant at all.
Nevertheless, $T\sim |m\rangle\langle m|$ can be regarded
as an approximate solution, since
both $aTa$ and $a^+Ta^+$ are slightly off-diagonal and
quite close to $NT$ and $TN$, and the R.H.S. of
\eomo\ almost cancel. 

The energy difference between the brane-anti-brane configuration
and a vortex on this system is given by
\eqn\env{\Delta \tr[Z,\bar{Z}]^2=-2\theta^2\tr |0\rangle\langle 0|
=-2\theta^2,}
thus indeed the vortex solution is a lower energy 
configuration. On the other hand, the approximate solution
$T=|m\rangle\langle m|$ carries more energy
\eqn\mener{\Delta \tr [Z,\bar{Z}]^2= 2(m-1)\theta^2.}
So the system will not tend to create these anisotropic
vortices. 

More vortex solutions can be simply generated by translating
$|0\rangle\langle 0|$ to other points, using the
matrix translation together with a gauge transformation
discussed in sect.2.

It would be interesting to consider the situation when both
gauge fields and the tachyon are turned on, and to look
for nontrivial solutions to $[Z, \bar{Z}]=0$. This will help
to understand the enigmatic nature of the nothing state 
\refs{\gmas, \sen}. Since the tachyon can never be gauged
away, as shown in the previous section, a nothing state involving
nonvanishing tachyon can not be equivalent to a pure
gauge field configuration. It is also interesting to see
how the $U(1)$ confinement problem is resolved in the present
context \refs{\piljin,\ashoke}.

In conclusion, we have found that the brane-anti-brane
system in matrix models has a rather strange representation.
This strangeness is due to the asymmetric treatment of
of the brane and the anti-brane, and the noncommutative
field theory is defined in favor of the brane. It is highly
desirable to find a new algebraic structure incorporating
different noncommutative structures on both branes. It
may help to follow the line of \chl\ to use a similarity
transformation to figure out this structure.

Acknowledgments. This work was supported by a grant of NSC and by a 
``Hundred People Project'' grant of Academia Sinica.

\vfill
\eject

\listrefs

\end